\documentclass[twocolumn,aps,floatfix,showpacs]{revtex4}
\newcommand{\be}{\begin{equation}}
\newcommand{\ee}{\end{equation}}
\newcommand{\bea}{\begin{eqnarray}}
\newcommand{\eea}{\end{eqnarray}}

\usepackage{graphicx}% Include figure files
\usepackage{bm}% bold math
\begin{document}

\title{Control of the interaction in a Fermi-Bose mixture}
\author{M. Zaccanti, C. D'Errico, F. Ferlaino, G.Roati, M. Inguscio, G. Modugno} \affiliation{LENS and
Dipartimento di Fisica, Universit\`a di Firenze,
  and INFM-CNR\\  Via Nello Carrara 1, 50019 Sesto Fiorentino, Italy}

\begin{abstract}
We control the interspecies interaction in a two-species atomic
quantum mixture by tuning the magnetic field at a Feshbach
resonance. The mixture is composed by fermionic $^{40}$K and bosonic
$^{87}$Rb. We observe effects of the large attractive and repulsive
interaction energy across the resonance, such as collapse or
a reduced spatial overlap of the mixture, and we accurately locate
the resonance position and width. Understanding and controlling instabilities
in this mixture opens the way to a variety of applications,
including formation of heteronuclear molecular quantum gases.
\end{abstract}
\pacs{03.75.Ss, 03.75.Kk, 34.20.Cf}
% Degenerate Fermi gases; dynamic properties of condensates; interatomic potentials and forces

\date{\today}
\maketitle Interactions between particles affect profoundly the
energy spectrum of ultracold quantum gases and determine most of
their properties, including superfluidity. Many different physical
scenarios have been opened by magnetically tunable Feshbach
resonances \cite{inouye}, which can be exploited to tune the
interactions beyond their natural values. The availability of such
resonances for most Bose and Fermi gases currently explored in
experiments has allowed to study a series of fundamental phenomena,
such as superfluidity in Fermi gases at the BEC-BCS crossover
\cite{becbcs}, formation of molecular quantum gases \cite{molecular
gases}, or observation of Efimov states \cite{kraemer}.
Heteronuclear mixtures and in particular mixtures of the two quantum
statistics, i.e. Fermi-Bose systems, represent an even richer
system, since more than one interaction between the constituent
particles come into play. In particular, achieving a control of the
interspecies interaction in these system is expected to give access
to a large range of new phenomena. Notable examples are dipolar
molecular gases \cite{fmgases}, boson-induced fermionic
superfluidity \cite{pairing}, novel quantum phases in strongly
correlated systems \cite{novelqp}. Feshbach resonances have been so
far studied in mixtures of $^{6}$Li-$^{23}$Na \cite{mit},
$^{40}$K-$^{87}$Rb \cite{simoni,jila,ferlaino}, $^{6}$Li-$^{7}$Li
\cite{ens}, but no fine tuning of the interaction has been reported
so far.

In this work we report a study of the properties of Fermi-Bose
mixture at an interspecies Feshbach resonance, which we exploit to
tune the fermion-boson scattering length $a_{FB}$ within a large
range of positive and negative values. We are able to drive  the
system into a regime where the interaction energy dominates the
behavior, leading to the onset of instabilities \cite{molmer}. In
particular we observe a collapse of the system for large attractive
interactions and the effects of a reduced spatial overlap of the components for large
repulsive interactions. We study the boundaries between regions
characterized by $a_{FB}$ with opposite sign to precisely determine
the resonance parameters that are necessary to realize a fine tuning
of the interaction. Fast tuning of $a_{FB}$ allows us to study the
dynamics of the system at collapse and to determine the timescale on
which experiments can be performed in the region of large negative
$a_{FB}$. This region is relevant for applications such as molecule
formation and boson-induced superfluidity \cite{pairing}.

\begin{figure}[ht]
\includegraphics[width=7.5cm]{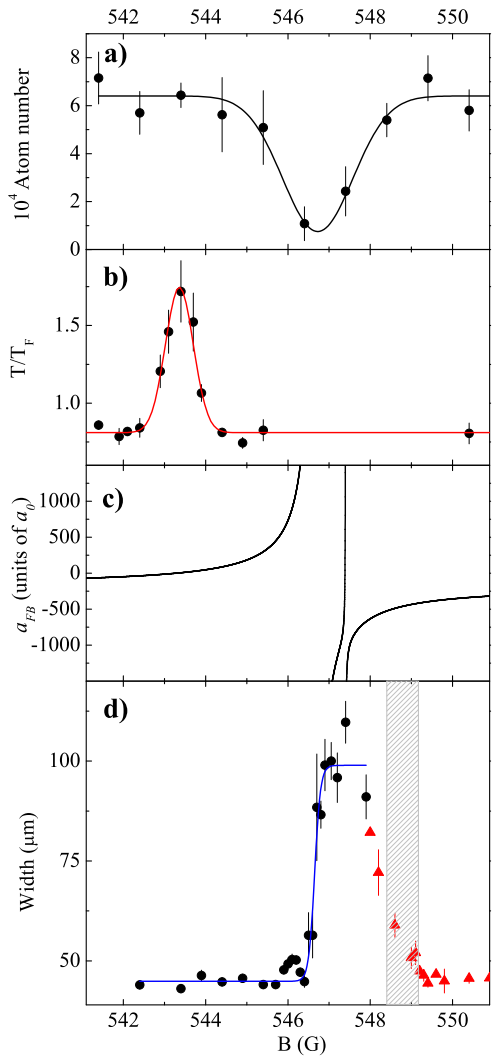}
\caption{(Color online) a) K atom number left in the trap after 50~ms hold time at
a fixed magnetic field $B$ in nondegenerate mixture
($T$=1.1~$\mu$K); b) efficiency of sympathetic cooling of fermions around the zero-crossing position;
c) theoretical expectation for the fermion-boson
scattering length $a_{FB}$ at the Feshbach resonance; d) width of
the Rb distribution after ballistic expansion from the trap. Here
the degenerate mixture is prepared at $B$=543.4~G (circles) or
$B$=551~G (triangles) and adiabatically brought to the final field
B. The highlighted region marks the transition from a stable to an
unstable regime.} \label{fig1}
\end{figure}

For the experiment we employ a $^{40}$K-$^{87}$Rb mixture, which is
produced using techniques already presented in detail elsewhere
\cite{roati,ferlaino}. We prepare samples of typically 10$^5$ K
fermions and 5$\times$10$^5$ Rb bosons at about 1$\mu$K in a
magnetic trap. The mixture is then transferred to an optical trap
created by two off-resonance laser beams, at a wavelength of
1030~nm, crossing in the horizontal plane. The trap depth for both
species is about 5~$\mu$K, and the trap frequencies $\omega/2\pi$
are (120,92,126)~Hz for Rb and a factor about $\sqrt{(87/40)}$
larger for K. The atoms are prepared in their absolute ground state
$|F=9/2, m_F=-9/2\rangle$ for K and $|1, 1\rangle$ for Rb. A
homogeneous magnetic field is then raised to $B$$\approx$550~G, in
the vicinity of the broadest K-Rb Feshbach resonance for these
states, shown in Fig. \ref{fig1} \cite{jila,ferlaino}. The samples
are then further cooled by reducing the depth of the optical trap in
2.4~s and then recompressed to the full depth in 150~ms
\cite{note1}. This allows to produce samples composed of up to
10$^5$ atoms per species, at $T<0.2 T_c$ for Rb and
$T$$\approx$\,0.3$T_F$ for K, where $T_c$=230~nK and $T_F$=630~nK.
The Bose gas is completely enclosed in the Fermi gas, whose
dimensions are approximately twice the ones of the BEC.
The Feshbach resonance we employ was so far studied only in thermal
samples \cite{jila,ferlaino}, where its presence is signalled by an
increase of three-body atom loss centered at $B$=546.7(4)~G
\cite{note2}, as shown in Fig.\ref{fig1}a. In a boson-fermion
mixture three-body processes involving two bosons and one fermion
are the dominant loss mechanism. They are predicted to depend on
bosons and fermions density distributions $n_B$ and $n_F$ and on the
interspecies scattering length as $\Gamma_3=K_3\int n_B^2n_Fd^3x$ ,
where $K_3\propto a_{FB}^4$ \cite{esry}. The maximum of the losses
therefore indicates the position of the resonance center $B_0$. In
addition to losses, we see also a heating of the system due to the
density dependence of $\Gamma_3$ that favors the loss of the coldest
atoms at the center of the distributions \cite{threebody}.

In order to have a precise control of the scattering length, it is
necessary to gain information on all parameters that determine the
resonance shape $a_{FB}(B)=a_{bg}(1-\Delta/(B-B_0))$. These are the
background scattering length $a_{bg}$, the center $B_0$ and the
width $\Delta$. The former has been determined through Feshbach
spectroscopy \cite{ferlaino,note2} to be $a_{bg}$=-185(4)~$a_0$ ,
and the expectation for the width is $\Delta\approx$-3~G. This
implies that in the vicinity of the resonance $a_{FB}$ is positive
only in a region extending from $B_0$ to about 3~G below, where
$a_{FB}$ crosses zero, as shown in Fig.\ref{fig1}c. Determination of
the boundaries between $B$-field regions with opposite signs of
$a_{FB}$ allows to determine experimentally the resonance
parameters. In particular, we locate the zero-crossing position by
studying the efficiency of sympathetic cooling of fermions during an
evaporation at magnetic fields below the resonance. The efficiency
is directly related to the interspecies thermalization rate, which
in the vicinity of the zero-crossing varies as $a_{FB}^2$. As shown
in Fig.\ref{fig1}b, the temperature of the fermionic component at
the end of the evaporation shows a marked increase centered at
543.4(2)~G, which we identify as the $a_{FB}$=0 position.

The other physical property of the system that is directly related
to $a_{FB}$ is the interspecies interaction energy
$U_{FB}=2\pi\hbar^2a_{FB}/\mu \int n_B n_Fd^3x$, where $\mu$ is the
reduced mass of the system. Each component is therefore felt by the
other one as an attractive (repulsive) potential for negative
(positive) $a_{FB}$. At a Feshbach resonance $U_{FB}$ is large and
it can substantially modify the distributions $n_B$, $n_F$. This can
lead to phenomena such as collapse for $a_{FB}<0$ or
phase separation for $a_{FB}>0$ \cite{molmer}, and will also
affect the behavior of three-body losses. We expect the resonance
center to be a sharp interface between these two opposite scenarios.

\begin{figure}[htbp]
\includegraphics[width=\columnwidth,clip]{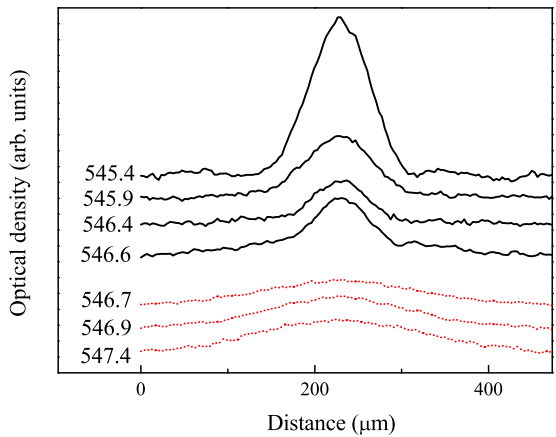}
\caption{(Color online) Profiles of the Bose gas in the mixture for various final
values of the magnetic field ramp starting at $B$=543.4~G. The
condensate survives despite the large atom losses as long as $B<B_0$
(continous lines). For $B>B_0$ the condensate is destroyed by
collapse (dotted lines). } \label{fig2}
\end{figure}

We have investigated these phenomena by adiabatically sweeping the
magnetic field across the Feshbach resonance. In a first experiment
the field is increased in 50~ms from $B_i$=543.4~G \cite{note} to a
final field $B_f$ that is varied from 543.4~G to 548~G. We
concentrate our attention on the bosonic component of the mixture,
since it has a lower chemical potential, and is therefore more
strongly affected by variations in $U_{FB}$. Fig.\ref{fig2} shows
the typical profiles of the Bose gas after 10~ms of permanence at
$B_f$ and 18~ms of expansion from the optical trap at $B\approx$0~G
\cite{note3}. We observe a decrease of the atom number as large as
80\% when $B_f$ approaches $B_0$, but the condensate surprisingly
survives. This observation is in contrast with the expected heating
associated to three-body atom losses, and is a strong indication that
the repulsive interaction is driving the system into a regime of phase separation.
The absence of heating indeed indicates that three-body losses remove
preferentially the warmest atoms in the system, and therefore that the only
overlap of the two clouds comes from their boundary regions,
where the most energetic atoms reside.
A mean-field model of our system \cite{modugnom} confirms that for the large positive
$a_{FB}$ expected for $B$ close to $B_0$ the two components tend to phase
separate into two vertically stacked domains, due to the anisotropy originated by gravity.

Once the field is tuned above $B_0$, the $U_{FB}$ suddenly changes
sign, and the two components tend to collapse at the center of the
trap because such energy is larger than the local intraspecies
repulsive energy \cite{collapse,hamburg}. The sudden increase of the
density overlap of the two components now promotes the loss of the
coldest atoms at the trap center, with a resulting rapid heating of
the system. We observe this collapse as $B_f$ is tuned above
546.6~G, where the condensate disappears, and one is left with a
thermal gas at $T\approx$600~nK. This study therefore indicates that
the scattering length changes sign between 546.6 and 546.7~G. This
value of $B_0$ is in good agreement with the value extracted from
the loss feature in a thermal mixture. Fig.\ref{fig1}d (circles)
summarizes this behavior. In this Figure we plot the vertical width
of the Bose gas when fitted with a single gaussian profile. This
shows the clear transition from the $a_{FB}>$0 regime at $B<B_0$
where the width stays approximately constant to the $a_{FB}<$0
regime where the condensate is destroyed by the collapse
instability. The Fermi gas behaves in a similar way, featuring a
sudden increase of the temperature when the field is tuned above
$B_0$. A phenomenological fit of the experimental data with a
Boltzmann growth function indicates $B_0$=546.65(20)~G.

Note that the width of the Bose gas
shown in Fig.\ref{fig1}d features a small local maximum around
546~G, where $a_{FB}$ is of the order of 1000~$a_0$, that we interpret as an
increased confinement felt by the bosons in the trap due to the
repulsion by the fermions. The width drops again in the vicinity of
the resonance because of the reduction in the atom numbers. Further
investigation of this regime of large positive $a_{FB}$ is
in progress. Fig.\ref{fig1}c shows also a very narrow spin resonance predicted by our
quantum collisional model \cite{ferlaino}. We actually detect
it in the experiment as an increase in the loss and heating rate in a
magnetic field range of 100~mG centered at 547.4(1)~G.

\begin{figure}[thbp]
\includegraphics[width=\columnwidth,clip]{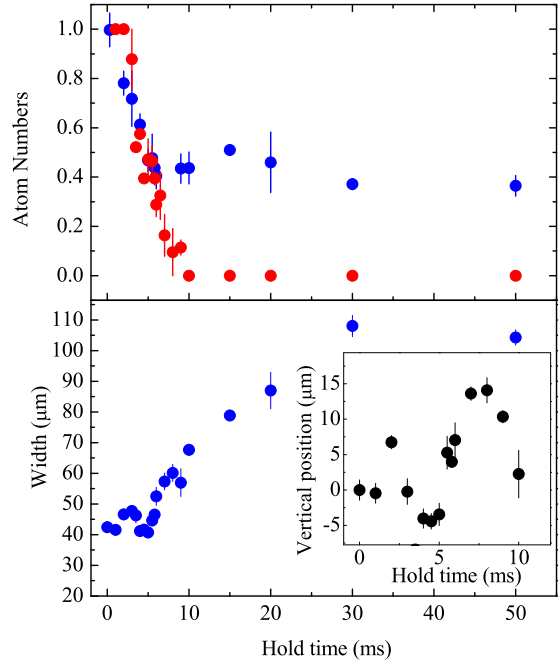}
\caption{(Color online) a) Evolution of the atom number in the bosonic
component(circles) and in the fermionic component (triangles) after
a rapid change of the magnetic field from $B_f$=551~G to
$B_f$=547.5~G. b) Evolution of the width and vertical position
(inset) of the bosonic component. } \label{fig3}
\end{figure}

The dynamics of collapse can now be studied in detail, thanks to the
possibility of tuning of $a_{FB}$ to arbitrarily large values in
times as short as a few 100~$\mu$s. Fig.\ref{fig3} shows the
time-evolution of a mixture that was prepared at $B_i\approx$551~G,
where $a\approx$-300~$a_0$, and then rapidly brought to
$B_f$=547.5~G where we expect a scattering length
$a_{FB}$=-700$_{-800}^{-600}$~$a_0$ that is too large for the
system to be stable. We observe a rapid loss of about 2/3 of the
atoms in both components in the first 5-6~ms, while both the width
and the center of the bosonic momentum distribution start to
oscillate at frequencies of the order of the bare trap frequency.
The oscillations are the evidence of the presence of a large
attractive interaction energy $U_{FB}$, which modifies the effective
trap potential experienced by the bosons. The mutual attraction
tends to increase the overlap of the two components and therefore
also increases the rate of three-body losses beyond the $a_{FB}^4$
dependence. After 6~ms most of the atoms in the center of the two
clouds are lost and the condensate is almost totally heated into a
thermal sample. The subsequent decrease of density leads to a rapid
decrease of $\Gamma_3$, almost stopping the atom losses.

We have observed this rapid loss of atoms followed
by a much slower decay and a heating of the system associated to collapse
for magnetic fields up to $B$=548.6~G, while for fields larger than $B$=549.4~G the mixture
is clearly stable. In the intermediate region the system shows a
moderate loss of atoms and shape excitations, presumably due to
the sudden variation of the interaction energy. This behavior is summarized
by the measurement reported in Fig.\ref{fig1}d (triangles). Here we have prepared the
mixture at $B_i$=551~G, swept the field in 50~ms to a variable $B_f$
on the negative-$a_{FB}$ side of the resonance, and then waited for additional
20~ms. The width of the condensate starts to increase around $B_f$=549.4~G
until around  $B_f$=548.6~G the condensate is totally heated into a thermal
sample as a consequence of collapse. The
transition region between stable and unstable conditions is highlighted.
The corresponding scattering length range is $a_{FB}$=-600$\div$-350~$a_0$ in
qualitative accordance with the prediction of a static mean field model
\cite{modugnom} of a critical scattering length $a_{c}$=-400~$a_0$
for the nominal atom numbers $N_B$=$N_F$=4$\times$10$^4$ in this
specific experiment.

It is important to note that the collapse does not apparently take
place on timescales much smaller than the trap period, as it can be
expected. For example, if the field is brought from $B_i$=551~G to
$B_f$=547.5~G in less than 2~ms and then back to 551~G in the same
time, no apparent atom loss or excitation of the condensate is
observed.

The possibility of a resonant control of $a_{FB}$ will allow to test
various aspects of strongly interacting Fermi-Bose systems, such as
the scaling law for $a_c$ with the atom density \cite{modugnom}, the
dynamics of the expansion \cite{expansion}, and theoretical issues
including the role of beyond-mean-field corrections \cite{beyond}
dynamical mean-field models \cite{adikhari} and finite-temperature
effects \cite{hu}.

In conclusion, we have studied the behavior of a $^{40}$K-$^{87}$Rb
Fermi-Bose mixture at an interspecies Feshbach resonance, whose
magnetic-field position and width have been precisely determined by
locating the critical values for the interspecies scattering length
$a_{FB}$. Besides an accurate investigation of mean-field effects,
this resonance can be employed for precisely tuning $a_{FB}$ for a
variety of applications. In particular, we are now investigating the
possibility of associating pairs of atoms into molecules using
magnetic field sweeps across the resonance from the $a_{FB}<0$ side,
in analogy with what performed in homonuclear systems
\cite{molecular gases}. Preliminary studies indicate that the system
can survive sweeps from $B>B_0$ to $B<B_0$ as slow as 2~ms/G without
collapsing, since the time spent in the region of large negative
$a_{FB}$ is too short. Interestingly, this ramp speed is still
smaller than the minimum ramp speed needed in homonuclear systems
\cite{hodby} to achieve the optimal atom-molecule conversion with
adiabatic sweeps. This indicates the feasibility of producing
heteronuclear fermionic molecules in this mixture. Instabilities can
be totally avoided by preparing the mixture in optical lattices \cite{esslinger,hamburgl}.
The tunable interspecies interaction
will help develop new strategies for the production and
control of atomic and molecular quantum gases in lattices.

We acknowledge contributions by A. Simoni and M. Modugno, and
critical reading of the manuscript by C. Fort. F.F. is also at
Institut f\"{u}r Experimental Physik, Universit\"{a}t Innsbruck,
Austria. This work was supported by MIUR, by EU under contracts
HPRICT1999-00111 and MEIF-CT-2004-009939, by Ente CRF, Firenze and
by CNISM, Progetti di Innesco 2005.

Related investigations have been reported by a group in Hamburg after completion
of this work \cite{hamburg2}.


\begin{thebibliography}{99}
\bibitem{inouye} S. Inoyue, {\it et al.},
%M. R. Andrews, J. Stenger, H.-J. Miesner, D.M. Stamper-Kurn, and W. Ketterle,
 Nature {\bf 392}, 151 (1998).
\bibitem{becbcs} C. A. Regal, M. Greiner, and D. S. Jin
Phys. Rev. Lett. {\bf 92}, 040403 (2004); C. Chin, {\it et al.}
%M.Bartenstein, A. Altmeyer, S. Riedl, S. Jochim, J. Hecker Denschlag,
%and R. Grimm
, Science {\bf 305}, 1128 (2004); T. Bourdel, {\it et al.}
%L.Khaykovich, J. Cubizolles, J. Zhang, F. Chevy, M. Teichmann, L.
%Tarruell, S. J. J. M. F. Kokkelmans, and C. Salomon
, Phys. Rev. Lett. {\bf 93}, 050401 (2004); G. B. Partridge, {\it et
al.}
%K. E. Strecker, R. I. Kamar, M. W. Jack, and R. G. Hulet
, Phys. Rev. Lett. {\bf 95}, 020404 (2005); M. W. Zwierlein, {\it et
al.}
% J. R. Abo-Shaeer, A. Schirotzek, C. H. Schunck, W. Ketterle
, Nature {\bf 435}, 1047 (2005).
\bibitem{molecular gases} S. Jochim, {\it et al.}
%M. Bartenstein, A. Altmeyer, G. Hendl, S. Riedl, C.
%Chin, J. Hecker Denschlag and R. Grimm
, Science {\bf 302}, 2101
(2003)\textit; M. Greiner, C.A. Regal, and D.S. Jin, Nature {\bf
426}, 537 (2003); M. W. Zwierlein, {\it et al.}
%C. A. Stan, C. H. Schunck, S. M. F. Raupach, S. Gupta, Z. Hadzibabic, and W. Ketterle
, Phys. Rev. Lett. {\bf 91}, 250401 (2003).
\bibitem{kraemer} T. Kraemer, {\it et al.}
%M. Mark, P. Waldburger, J. G. Danzl, C. Chin, B. Engeser, A. D. Lange, K. Pilch, A. Jaakkola, H.-C. Naegerl, R. Grimm
, Nature {\bf 440}, 315
(2006).
\bibitem{fmgases} M. A. Baranov, M. S. Marenko, V. S. Rychkov, and G.V.
Shlyapnikov, Phys. Rev. A {\bf 66}, 013606 (2002); B. Damski, {\it
et al.},
%L. Santos, E. Tiemann, M. Lewenstein, S. Kotochigova, P. Julienne, and P. Zoller,
 Phys. Rev. Lett. {\bf 90}, 110401 (2003);
\bibitem{pairing} H. Heiselberg, C. J. Pethick, H. Smith, and L. Viverit,
 Phys. Rev. Lett. {\bf 85}, 2418 (2000); M. J. Bijlsma, A.
Heringa, and H.T. C. Stoof, Phys. Rev. A {\bf 61}, 052601 (2000); L.
Viverit, Phys. Rev. A {\bf 66}, 023605 (2002); D. V. Efremov and L.
Viverit, Phys. Rev. B {\bf 65}, 134519 (2002); F. Matera, Phys. Rev.
A {\bf 68}, 043624 (2003).
\bibitem{novelqp} A. Albus, F. Illuminati,
and J. Eisert Phys. Rev. A {\bf 68}, 023606 (2003); H.P.
B\"{u}chler, G. Blatter, and W. Zwerger, Phys. Rev. Lett. {\bf 90},
130401 (2003); M. Lewenstein, L. Santos, M. A. Baranov, and H.
Fehrmann, Phys. Rev. Lett. {\bf 92}, 050401 (2004).
\bibitem{mit} C. A. Stan, {\it et al.}
%M.W. Zwierlein, C. H. Schunck, S.M. F. Raupach, and W. Ketterle
, Phys. Rev. Lett. {\bf 93}, 143001 (2004).
\bibitem{simoni} A.~Simoni, {\it et al.}
%F.~Ferlaino, G.~Roati, G.~Modugno, and M.~Inguscio,
, Phys. Rev.
Lett. {\bf 90}, 163202 (2003).
\bibitem{jila} S. Inouye, {\it et al.}
%J. Goldwin, M. L. Olsen, C. Ticknor, J. L. Bohn, and D. S. Jin
, Phys. Rev. Lett. {\bf 93}, 183201 (2004).
\bibitem{ferlaino} F. Ferlaino, {\it et al.}
% C. D'Errico, G. Roati, M. Zaccanti, M. Inguscio, G. Modugno, A. Simoni
, Phys. Rev. A  {\bf 73}, 040702
(2006).
\bibitem{ens} J. Zhang, {\it et al.}
%E. G. M. van Kempen, T. Bourdel, L. Khaykovich, J. Cubizolles, F.
%Chevy, M. Teichmann, L. Tarruel, S. J. J. M. F. Kokkelman, and C.Salomon
, in Proceedings of the XIX International Conference on
Atomic Physics, L. G. Marcassa, V. S. Bagnato, K. Helmerson eds.
(AIP, New York, 2005).
\bibitem{molmer} K. Molmer, Phys. Rev. Lett. {\bf 80}, 1804 (1998).
\bibitem{roati} G. Roati, F. Riboli, G. Modugno, and M.
Inguscio, Phys. Rev. Lett. {\bf 89}, 150403 (2002).
\bibitem{note1} Selective evaporation of bosons is achieved by exploiting the
lower trap depth for bosons due to the larger contribution of
gravity in shallow traps.
\bibitem{note2} We have corrected a minor magnetic-field
calibration error committed in \cite{ferlaino}. This results in a
positive shift of the $B_0$ by 1.4~G, and in a corresponding shift
of $a_{bg}$ by 8~$a_0$.
\bibitem{esry} J. P. D'Incao and B. D. Esry, Phys. Rev. A {\bf 73},
030702(R) (2006).
\bibitem{threebody} T. Weber, J. Herbig, M. Mark, H.-C. N\"{a}gerl, and R. Grimm,
Phys. Rev. Lett. {\bf 91}, 123201 (2003).
\bibitem{note} The mixture is evaporated at $B=538 G$ and then
brought adiabatically to the zero-crossing.
\bibitem{note3} We switch off the magnetic field to reduce $a_{FB}$ to $a_{bg}$ during
the first phases of the expansion \cite{expansion}.
\bibitem{modugnom} M. Modugno, {\it et al.}
%F. Ferlaino, F. Riboli, G. Roati, G. Modugno, and M. Inguscio
, Phys. Rev. A  {\bf 68}, 043626 (2003).
\bibitem{collapse} G.~Modugno, {\it et al.}
%G.~Roati, F.~Riboli, F.~Ferlaino, R.~J.~Brecha, and M.~Inguscio
, Science {\bf 297}, 2200 (2002).
\bibitem{hamburg} C. Ospelkaus, S. Ospelkaus, K. Sengstock and K. Bongs, Phys. Rev. Lett.  {\bf 96},
020401 (2004).
\bibitem{expansion} F. Ferlaino, {\it et al.}
%E. de Mirandes, G. Roati, G. Modugno, and M. Inguscio
, Phys. Rev. Lett.  {\bf 92}, 140405
(2004).
\bibitem{beyond} A. P. Albus, F. Illuminati, and M. Wilkens, Phys. Rev. A {\bf 67},
063606 (2003).
\bibitem{adikhari} S. K. Adhikari, Phys. Rev. A {\bf 70} 043617 (2004).
\bibitem{hu} X.-J. Liu, M. Modugno and H. Hu, Phys. Rev. A {\bf 68},
053605 (2003).
\bibitem{hodby} E. Hodby, {\it et al.}
%S. T. Thompson, C. A. Regal, M. Greiner, A. C. Wilson, D. S. Jin, E.
%A. Cornell, and C. E. Wieman
, Phys. Rev. Lett. {\bf 94}, 120402 (2005).
\bibitem{esslinger} K. G\"{u}nther, {\it et al.}
% T. Sth\"{o}ferle, H. Moritz, M. K\"{o}hl, T. Esslinger
, Phys. Rev. Lett. {\bf 96}, 180402
(2006).
\bibitem{hamburgl} S. Ospelkaus, {\it et al.}, Phys. Rev. Lett. {\bf 96}, 180403
(2006).
\bibitem{hamburg2} C. Ospelkaus, {\it et al.}, Phys. Rev. Lett. {\bf 97}, 120402
(2006); S. Ospelkaus, {\it et al.}, {\it ibid.} {\bf 97}, 120403
(2006).

\end{thebibliography}
\end{document}